\documentclass{article}
\usepackage{spconf,amsmath,graphicx}
\usepackage{booktabs}
\usepackage{amsfonts}
\usepackage{multirow}
\usepackage{tablefootnote}
\usepackage{hyperref}
\hypersetup{
    colorlinks=true,
    anchorcolor=black,
    citecolor=black,
    linkcolor=black,
    filecolor=black,      
    urlcolor=black,
    pdftitle={Overleaf Example},
    pdfpagemode=FullScreen,
    }

\title{Self-Supervised Hierarchical Metrical Structure Modeling}
\name{Junyan Jiang$^{1, 2}$ \hspace{1cm} Gus Xia$^{1, 2}$}
\address{$^1$ Music X Lab, NYU Shanghai \hspace{1cm}$^2$ Mohamed bin Zayed University of Artificial Intelligence \hspace{1cm}}
\begin{document}
\ninept
\maketitle
\begin{abstract}

We propose a novel method to model hierarchical metrical structures for both symbolic music and audio signals in a self-supervised manner with minimal domain knowledge. The model trains and inferences on beat-aligned music signals and predicts an 8-layer hierarchical metrical tree from beat, measure to the section level. The training procedure does not require any hierarchical metrical labeling except for beats, purely relying on the nature of metrical regularity and inter-voice consistency as inductive biases. We show in experiments that the method achieves comparable performance with supervised baselines on multiple metrical structure analysis tasks on both symbolic music and audio signals. All demos, source code and pre-trained models are publicly available on GitHub\footnote{\href{https://github.com/music-x-lab/Self-Supervised-Metrical-Structure}{https://github.com/music-x-lab/Self-Supervised-Metrical-Structure}}.
\end{abstract}
\begin{keywords}
Metrical structure, self-supervised learning, music understanding, hierarchical structure analysis
\end{keywords}
\section{Introduction}
\label{sec:intro}

Metrical structure analysis is a useful task for music information retrieval (MIR) \cite{maddage2004content, ellis2007identifyingcover}, computer musicology \cite{dixon2004towards} and music generation \cite{dai2022missing}. With the development of deep learning, we have witnessed improved performance on beat and downbeat tracking \cite{bock2019multi,bock2016joint, zhao2022beat}, part segmentation \cite{mccallum2019unsupervised} and hierarchical metrical structure analysis \cite{bock2016joint, zhao2022beat, supervised}. However, most of these methods are based on supervised training, which requires manual ground-truth labeling of a large music collection. This poses two potential risks: (1) the labeling procedure is time-consuming and requires domain experts; (2) the ambiguous nature of music and the subjectivity of human annotators commonly cause biases in MIR datasets \cite{koops2020automatic}.

Self-supervised learning has been a promising way to model signals without labeled datasets \cite{bert, misra2020self, manocha2021cdpam}. Self-supervised learning is recently introduced to MIR tasks including pitch estimation \cite{spice}, music classification \cite{zhao2022s3t}, cover song identification \cite{yao2022contrastive} and tempo estimation \cite{quinton2022equivariant}. These works inspire us to utilize self-supervised learning in hierarchical metrical structure analysis, where the scarcity of labeled datasets limits the model's performance.

\begin{figure}[t]
    \centering
    \includegraphics[width=\linewidth,clip, trim=1.7cm 2.7cm 8.5cm 0.4cm, page=3]{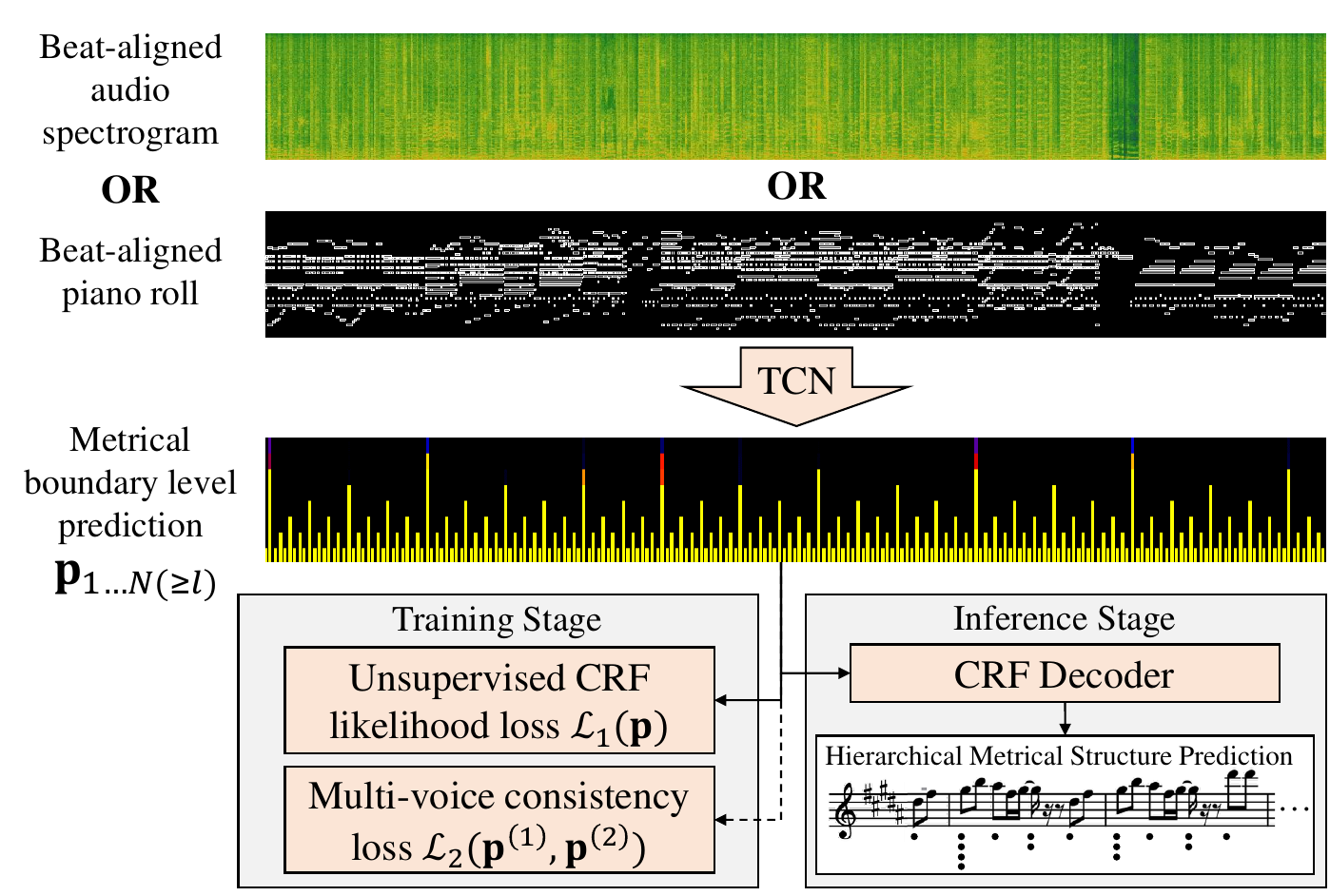}
    \caption{An overview of the proposed method. The model accepts beat-aligned spectrogram or piano roll as inputs and predicts a hierarchical metrical tree of 8 layers beyond tatums. The model trains under self-supervision, i.e., no ground truth of $\mathbf{p}$ is used in training.}
    \label{fig:overview}
\end{figure}

Hierarchical metrical structure modeling aims to describe temporal rhythmic patterns of music at different time scales. Different levels of metrical structures are interrelated (e.g., multiple beats form a measure), building up the metrical hierarchy. The concept can be further extended to larger time scales, forming concepts like hypermeasures and hypermeters \cite{temperley2008hypermetrical, gttm}, which are useful in analyzing other high-level features like melodic phrase structure, harmonic structure and form \cite{robins2017phrase}. The notation of hierarchical structure is formally described in the Generative Theory of Tonal Music (GTTM) \cite{gttm} by using different numbers of dots to represent different metrical boundary levels (as in fig.\ \ref{fig:overview} lower right).

In this paper, we describe a self-supervised approach to hierarchical metrical structure modeling (see fig.\ \ref{fig:overview}). A Temporal Convolutional Network (TCN) accepts a beat-aligned spectrogram or piano roll as input and predicts a metrical boundary level for each time step. For simplicity, we ignore non-binary metrical structures (e.g., 3/4 or 6/8 meters) and focus on binary structures in the paper (temporary binary irregularity is still allowed).

We encourage the model to predict a sequence that follows binary regularity (i.e., a level-$l$ metrical unit contains exactly 2 level-$(l-1)$ metrical units). We implement this inductive bias using an unsupervised Conditional Random Fields (CRF) loss. We show that by minimizing this loss term, the model learns a semantically meaningful hierarchical metrical structure up to some global offset. Similar to other self-supervised approaches \cite{spice, quinton2022equivariant}, the global offset is shared across all songs and can be calibrated with a few annotated samples (1 song), which is the only supervised data required for the model. For multi-track data like MIDI files, we also propose a consistency-based loss that further boosts the model's performance.

The paper's main contributions are:
\begin{itemize}\itemsep0em 
    \item We provide a novel self-supervised approach for hierarchical metrical structure analysis. The model performance is comparable with supervised methods on several tasks.
    \item We propose an unsupervised training target based on metrical regularity, and show its effectiveness on metrical structure modeling on multiple scales.
    \item As far as we know, our model is the first hypermetrical structure analyzer for audio signals, opening up new possibilities for downstream MIR applications.

\end{itemize}


\section{Methodology}
\label{sec:method}

\subsection{Design Intuition}

Before discussing the details of the model, we first describe the intuition behind the self-supervision setting. One common way to train self-supervised models is to introduce pseudo-labels. In contrastive learning, several views of a same data sample form positive pairs, and views of different samples form negative pairs \cite{manocha2021cdpam,jaiswal2020survey}. An equivalent-based data augmentation also creates pseudo-labels with relative distance constraints, preventing the model from trivial solutions like predicting everything as a single label \cite{spice,quinton2022equivariant}. 

\begin{figure}[t]
    \centering
    \includegraphics[width=\linewidth,clip, trim=0.2cm 13.4cm 16.5cm 0.2cm, page=10]{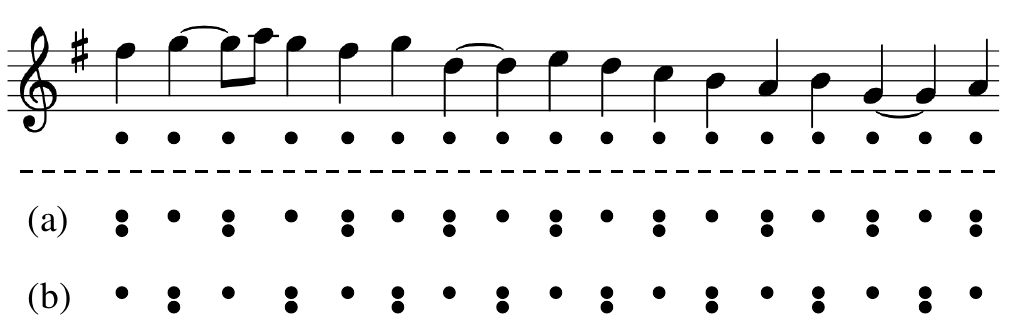}
    \caption{The intuition of self-supervision for metrical analysis. For simplicity, assuming the level-$(l-1)$ metrical boundaries are known and we want to build the level-$l$ metrical structure. Under the assumption of binary regularity, we strongly prefer one of the choices in (a) and (b). In either case, the labeling for consecutive beats is different, forming negative sample pairs in the sense of contrastive methods.}
    \label{fig:intuition}
\end{figure}

In self-supervised metrical analysis, the main source of supervision comes from the preference of metrical regularity. For binary metrical regularity (with a periodicity of 2), a level-$l$ metrical always contains 2 level-$(l-1)$ metrical units, forming an alternating pattern as shown in fig.\ \ref{fig:intuition}. To relate with contrastive methods, we can regard two consecutive level-$(l-1)$ measure boundaries as negative pairs for level-$l$ boundary prediction and two level-$(l-1)$ measure boundaries with distance 2 as positive pairs. Instead of using a constrastive loss which regards metrical regularity as hard constraints, we design an unsupervised CRF loss to  incorporate it as a soft constraint with controllable degree of penalty to metrical irregularity. Also, we design our CRF loss to enable joint estimation of multiple metrical levels instead of just a single level (see section \ref{sec:crf}).




\subsection{Model Architecture}

The backbone of our model is 6 stacked temporal convolutional layers with the same configuration as \cite{supervised}. For symbolic input (MIDI file) with $T$ tracks, we regard the 16th-note-quantized (tatum-quantized) piano roll of each track $t$ as input $\mathbf{m}_t$. The TCN then outputs a metrical level prediction $\mathbf{h}_i^{(t)}$ and a confidence score $\alpha_i^{(t)}$ for each time step $i\in \{1,...,N\}$. Similar to \cite{supervised}, the final prediction is acquired by a weighted pooling over all tracks:
\begin{equation} \small
    a^{(t)}_i:=\exp{ \alpha^{(t)}_i} \big/ \sum_{t'}  { \exp{\alpha^{(t')}_i}}
\end{equation}
\vspace{-0.2 cm}
\begin{equation} \small
    \mathbf{p}_i:=\sum_t a^{(t)}_i \mathbf{p}^{(t)}_i=\sum_t a^{(t)}_i \textrm{Softmax}(\mathbf{h}^{(t)}_i)
\end{equation}
where $p_{il}$ denotes the predicted probability that time step $i$ has a metrical boundary of level $l \in \{0,...,L\}$. For convenience, we use $p_{i(\geq l)}=\sum_{l'=l}^L p_{il'}$ to denote the probability that the time step $i$ has a metrical boundary of at least level $l$.

For audio input, some changes are made compared to symbolic input. We use a Short-Time Fourier Transform (STFT) spectrogram instead of the piano roll as input, and the spectrogram is re-quantized to a 128th-note level using time stretching. We prepend 3 extra convolutional-pooling layers before the TCN. Each pooling layer has a kernel size of 2, bringing down the resolution to the 16th-note level to align with the time step of symbolic input. The extra convolutional layers are identical to the ones in TCN but with a constant dilation 1. Also, since we cannot acquire cleanly separated tracks of audio, the TCN directly predicts $\mathbf{h}_i$ without the weighted pooling operation, and the probability is calculated as $\mathbf{p}_i = \textrm{Softmax}(\mathbf{h}_i)$.

\subsection{Unsupervised CRF Loss} \label{sec:crf}

\begin{figure}[t]
    \centering
    \includegraphics[width=\linewidth,clip, trim=1.5cm 8.3cm 11.4cm 0.7cm, page=5]{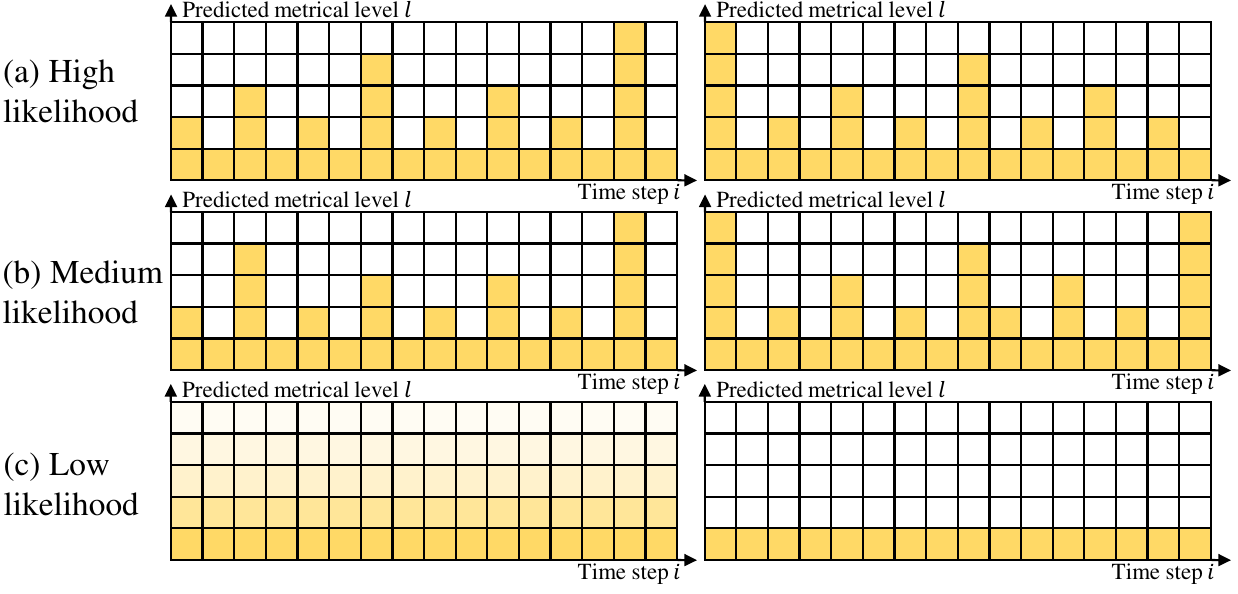}
    \caption{Examples of the unsupervised CRF likelihood. Each cell $(i, l)$ visualizes the predicted probability of a metrical level $p_{i(\geq l)}$ for $l=0...L$. Darker color denotes a higher value. (a) The likelihood is high since binary regularity is satisfied. (b) The likelihood is not as high since the binary regularity is violated with a level-2 insertion  (left) or a level-0 deletion (right). (c) The likelihood is very low for trivial predictions like assigning a uniform probability to all levels (left) or always predicting a single level (right).}
    \label{fig:crf_loss}
\end{figure}

The aim of the CRF loss is to encourage the TCN's output to be close to a sequence that follows binary regularity on all levels (see some examples in fig.\ \ref{fig:crf_loss}). The design of CRF's state space and transition potential is similar to \cite{supervised}, and we recommend the reader refer to it for detailed explanation and some concrete examples. We here adopt a simplified version to reduce computational cost. For a hierarchical metical structure with $L$ layers, the joint state space is defined as $z_i=(z_i^{(1)}, ..., z_i^{(L)})$ where each $z_i^{(l)}\in\{0, 1\}$ represents the the number of complete level-$(l-1)$ metrical units in this level-$l$ unit up to the current time step. If 2 (or more) complete level-$(l-1)$ metrical units are spanned, a new level-$l$ unit must start, otherwise a binary irregularity penalty is posed and we keep $z_i^{(l)}=1$. The transition potential is defined the same way as in \cite{supervised}:
\begin{equation} \small
    \phi(z_{i-1}, z_i)=\prod_{l=1}^{L}\left\{\begin{matrix}
    A^{(l)}_{z_{i-1}^{(l)}z_{i}^{(l)}} & l \leq l_i + 1 \\
    \mathbb{I}[z_{i-1}^{(l)}=z_{i}^{(l)}] &  l > l_i + 1
    \end{matrix}\right.
\end{equation}
where $l_i=\arg \max_{l=0...L}[z_i^{(1...l)}=0]$ denotes the corresponding metrical boundary level of $z_i$. If $z_i^{(1)}\neq 0$, we have $l_i=0$, i.e., the lowest metrical boundary level (tatum-level) only. The level-$0$ states $z_i^{(0)}$ transits every time step since $l\leq l_i + 1$ always holds for $l=0$. For $l\geq 1$, the state of each level $z_i^{(l)}$ transits if and only if a level-$(l-1)$ boundary is encountered ($l_i\geq l - 1$), otherwise the level-$l$ state must keep unchanged as the previous one, i.e., $z_{i-1}^{(l)}=z_{i}^{(l)}$. $\mathbf{A}^{(l)}$ is the single-layer transition potential matrix for level-$l$ metrical boundaries:
\begin{equation} \small
    \mathbf{A}^{(l)} = \begin{bmatrix}
\exp{\big(-w_\mathrm{del}^{(l)}\big)} & 1\\
1 & \exp{\big(-w_\mathrm{ins}^{(l)}\big)}
\end{bmatrix}
\end{equation}
where $w_\mathrm{del}^{(l)} > 0, w_\mathrm{ins}^{(l)} > 0$ are hyperparameters that controls the penalty of a level-$l$ metrical unit deletion and insertion respectively. Setting both $w_\mathrm{del}^{(l)}, w_\mathrm{ins}^{(l)}$ to $+\infty$ will result in a hard binary regularity constraint (i.e., the states on level-$l$ must alternate between 0 and 1). Different from \cite{supervised}, the insertion transition is now a self-transition, thus can be applied several times in a larger metrical unit. The emission potential function is defined as $\psi(z_i, \mathbf{p}_i)=p_{il_i}$.

The unsupervised CRF loss is defined as the negative log likelihood of the observation sequence $\mathbf{p}_{1...N}$. This can be calculated by summing up all possible latent state sequences $\mathbf{z}$ over the unnormalized CRF likelihood:
\begin{equation} \small
    \mathcal{L}_1 (\mathbf{p}) = -\log \sum_{\mathbf{z}}\prod_{i=2}^N \phi(z_{i - 1}, z_{i}) \prod_{i=1}^N\psi(z_i, \mathbf{p}_i)
\end{equation}
which can be computed efficiently by the well-known dynamic-programming-based forward algorithm.


\subsection{Consistency Loss}

For metrical analysis of multi-track data, we also include the consistency loss $\mathcal{L}_2$. For two time-aligned track-wise prediction $\mathbf{p}^{(1)},\mathbf{p}^{(2)}$, we want them to indicate the same metrical structures. We achieve this by requiring the same CRF hidden states to emit both observations by defining the joint emission potential $\psi'(z_i, \mathbf{p}^{(1)}_i,\mathbf{p}^{(2)}_i)=p^{(1)}_{il_i}p^{(2)}_{il_i}$. The consistency loss is defined by:
\begin{equation} \small
    \mathcal{L}_2 (\mathbf{p}^{(1)}, \mathbf{p}^{(2)}) = -\log \sum_{\mathbf{z}}\prod_{i=2}^N \phi(z_{i - 1}, z_{i})\prod_{i=1}^N\psi'(z_i, \mathbf{p}_i^{(1)}, \mathbf{p}_i^{(2)})
\end{equation}

\subsection{Post-Training Calibration}

Since the model solely relies on self-supervision, the model's output might mismatch the true metrical boundaries by a global offset. For a well-trained model, imagine that if all model predictions shift left or right some constant time steps, both loss terms remain low. The global offset always lies in the receptive field of TCN, usually within $\pm$32 time steps. Therefore, we need to look into a few annotated samples to calibrate the global offset (in practice, 1 song is enough). The calibration only needs to be done once per trained model and no more annotated data is required before or after the procedure.

\section{Experiments}
\label{sec:experiment}

\begin{figure}[t]
    \centering
    \includegraphics[width=\linewidth,clip, trim=0.5cm 10.5cm 19.6cm 0.5cm, page=6]{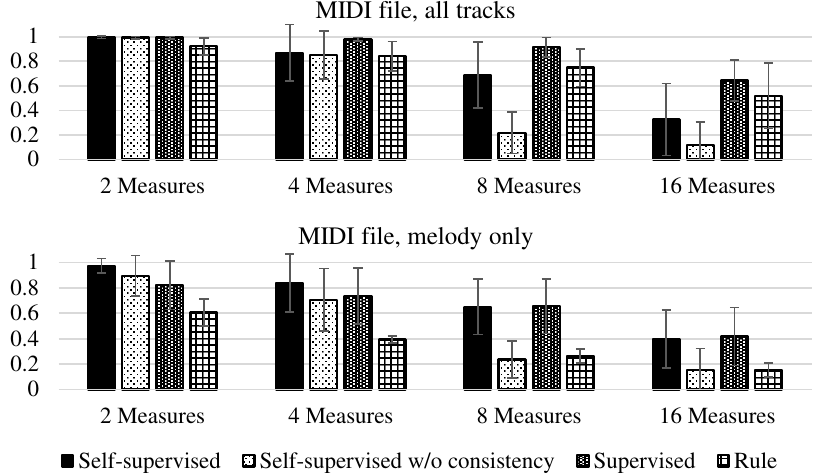}
    \caption{Evaluated mean and standard deviation of the F1 scores of hypermetrical structure analysis on the test split of the RWC-POP MIDI dataset. ``Melody only'' denotes a hard version of the task where only the melody track is used for prediction.}
    \label{fig:results_midi}
\end{figure}

\begin{figure}[t]
    \centering
    \includegraphics[width=\linewidth,clip, trim=0.5cm 15.3cm 19.6cm 0.4cm, page=7]{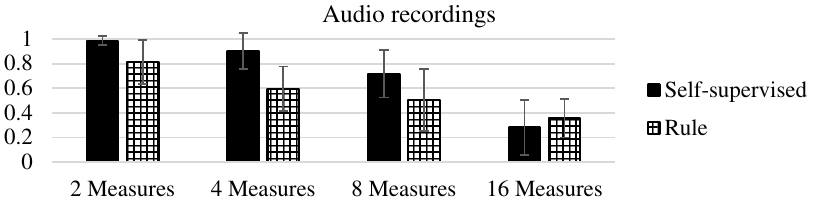}
    \caption{Evaluated mean and standard deviation of the F1 scores of hypermetrical structure analysis on the test split of the RWC-POP audio recordings. }
    \label{fig:results_audio}
\end{figure}

\subsection{Datasets}

We use the Lakh MIDI dataset \cite{raffel2016learning} to train the model for symbolic music analysis. We first acquire a subset of 3,739 multi-track MIDI files according to the criteria described in \cite{supervised} to exclude some low-quality files. We use the \texttt{pretty\_midi} package \cite{pretty_midi} to acquire beats and quantize each song into 16-note units. No hierarchical metrical information (e.g., downbeat annotations) is used in training. For evaluation, we use the RWC-POP MIDI dataset. We only use the test split in \cite{supervised} to make a fair comparison with the supervised model. A single song (RWC-POP-001, not in the test split) is used to calibrate the model.

To train the audio model, we collect a training set of 7,500 songs from an online music collection from a rhythm game \textit{osu!}\footnote{\href{https://osu.ppy.sh/beatmapsets}{https://osu.ppy.sh/beatmapsets}}. In this collection, every song comes with highly precise beat annotation by crowd annotators as beat labels are essential for rhythm game mapping. Downbeat annotations are also available, but they are not used in training. The collection has a main genre of J-pop but also contains a variety of other genres including western pop and rock. All audio files and annotations are publicly accessible, making it easier for reproduction and further research. The reader may refer to the paper's GitHub repository for more information about the dataset. For evaluation, we use the audio recordings of the RWC-POP dataset. We use the same data split as the symbolic model evaluation and the recording version of RWC-POP-001 to calibrate the model.

\begin{figure}[t]
    \centering
    \includegraphics[width=\linewidth,clip, trim=0.2cm 6.55cm 12.0cm 0.0cm, page=8]{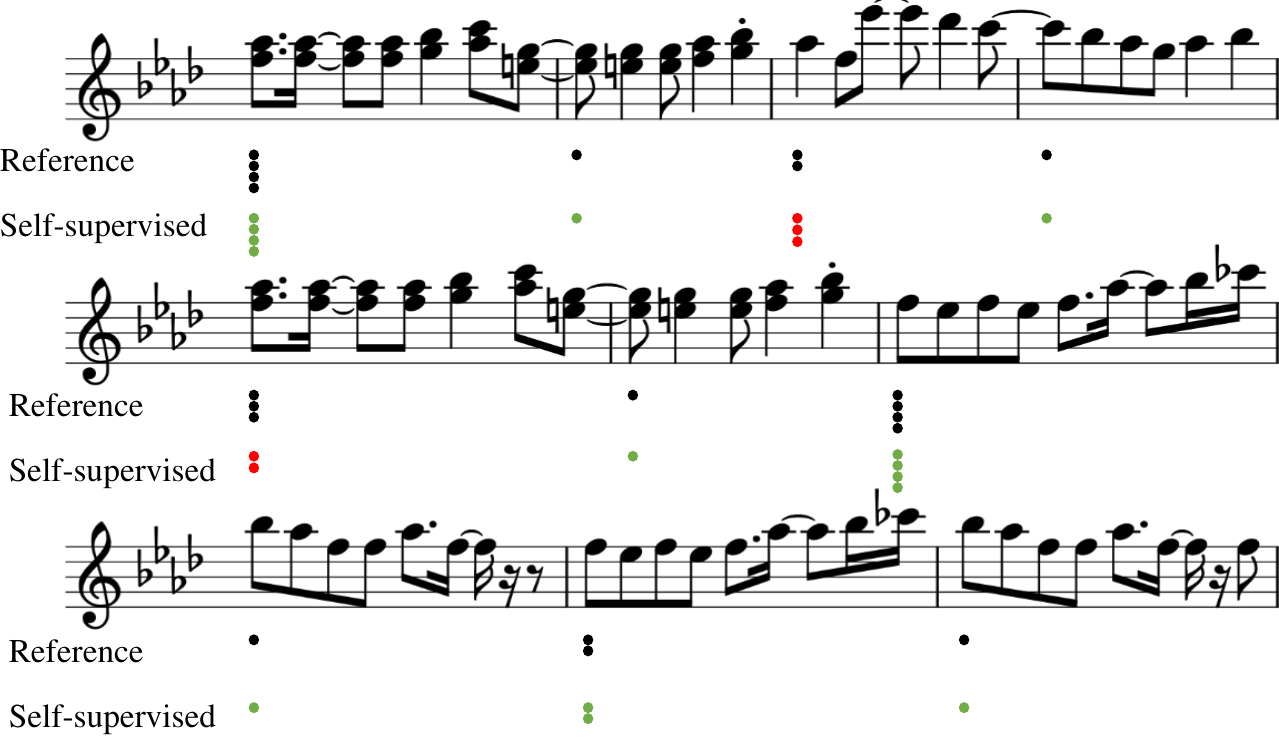}
    \caption{A case study of the symbolic hypermetrical structure modeling. The 10-measure piece (from RWC-POP-005) contains binary irregularity on a 4-measure scale (4+2+4). The model successfully identifies a temporary hypermeter change but deletes the wrong hypermeasure (2+4+4), resulting in wrong boundaries (marked in red). }
    \label{fig:case}
\end{figure}

\subsection{Model Hyperparameters}

We let the model predict a hierarchy of $L=8$ metrical layers from the 16th-note level. Since each level increases the metrical unit length by a factor of 2, the 2nd layer corresponds to beats and the 4th layer corresponds to measures for 4/4 songs. For both symbolic and audio music, we set $w_\mathrm{del}^{(l)}, w_\mathrm{ins}^{(l)}=+\infty$ for $l=1...4$ to disallow any binary irregularity up to a measure. We allow hypermeter transition by setting $w_\mathrm{del}^{(l)}=15$ and $w_\mathrm{ins}^{(l)}=20$ for $l=5...8$.

The format of the piano roll input aligns with \cite{supervised}. For audio input, we resample the audio to 22050Hz. Then, we use a 1024-point STFT with 75\% overlapping hann windows before time-stretching the spectrogram to a 128th-note level. Each model is trained for 10 epochs using Adam optimizer \cite{kingma2014adam} with a fixed $10^{-4}$ learning rate.

\subsection{Main Results: Hypermetrical Structure Prediction}

We first evaluate our model on metrical structure prediction beyond measures ($l=5...8$). For symbolic music analysis, we include 2 baseline models, the supervised model from \cite{supervised} trained on 50 annotated songs (726 MIDI tracks) and a rule-based model based on homogeneity metrics on different scales \cite{Serr2012UnsupervisedDO}. To make a fair comparison, the prediction of the self-supervised model will go through the same CRF decoder with the same hyperparameters as the supervised model. In this experiment, the downbeat labels are assumed to be known at test time; only predictions on ground-truth downbeats are used for decoding hypermetrical structures for all models.

For symbolic hypermetrical structure evaluation, we also include an ablation test that removes the track-wise consistency loss. The results are shown in fig.\ \ref{fig:results_midi}. The self-supervised model shows impressive performance on 2-measure and 4-measure hypermetrical structures, and the consistency loss further boosts the performance. When only the melody track is used (melody only), the model even outperforms the supervised baseline, showing that a large unlabeled dataset is helpful for the model to understand metrical structures better from limited information.

For audio hypermetrical structure evaluation, the results are shown in fig.\ \ref{fig:results_audio}. Since we are unable to train a supervised baseline without overfitting an extremely limited training set (50 songs), the only viable way is to use self-supervised learning to gain reasonable performance for the task. Compared to MIDI files, audio hypermetrical structure analysis is more challenging since we cannot impose track-wise consistency loss, and the input feature is often noisier.

Notice that the prediction of the highest layer (the 16-measure level, or $l=8$) is poor for self-supervised models. The main reason is that a stable binary rhythmic structure on a 16-measure level is much less common compared to lower levels, invalidating our model assumption. Therefore, the model fails to learn meaningful representation with self-supervision.

We also observe that the proposed method does not perform well near hypermetrical changes, with an example shown in fig.\ \ref{fig:case}. The model is capable of detecting most hypermetrical changes, but the labeling of the transition part is sometimes not musically meaningful. The main reason is that the preferences of irregular hypermeasure decomposition is not encoded into our CRF loss, and the model chooses the wrong decomposition under self-supervision.


\subsection{Downbeat Detection}

\begin{table}
\centering
\begin{tabular}{|c|c|c|c|}
\hline
\textbf{Input}                                                        & \textbf{Time Step}         & \textbf{Method}                                             & \textbf{Downbeat F1} \\ \hline
\multirow{4}{*}{\begin{tabular}[c]{@{}c@{}}Mel.\\ only\end{tabular}}  & \multirow{4}{*}{Tatum} & Proposed   & 87.67 $\pm$ 12.79   \\ \cline{3-4} &           & W/o consistency   & 86.18 $\pm$ 12.95    \\ \cline{3-4} & & Supervised \cite{supervised}   & 85.88 $\pm$ 13.64                \\ \cline{3-4}  &                        & Interactive GTTM \cite{tojo2013computational} & 43.46 $\pm$ 45.13   \\ \hline
\multirow{3}{*}{\begin{tabular}[c]{@{}c@{}}All\\ tracks\end{tabular}} & \multirow{3}{*}{Tatum} & Proposed  & 98.36 $\pm$ 3.25  \\ \cline{3-4} & & W/o consistency  & 97.66 $\pm$ 5.46  \\ \cline{3-4}  &  & Supervised \cite{supervised}  & 97.44 $\pm$ 6.78  \\ \cline{2-4}  \hline
\multirow{2}{*}{\begin{tabular}[c]{@{}c@{}}Synth.\\ audio\end{tabular}} &\multirow{2}{*}{Frame} & Madmom \cite{bock2016joint}  & 92.47 $\pm$ 16.41   \\ \cline{3-4}   &   &  Transformer \cite{zhao2022beat}\tablefootnote{\label{note:jingwei}Cross-validation results on RWC-POP synthesized audio and recordings.}   & 93.63 $\pm$ 12.26 \\ \hline
\end{tabular}
\caption{Evaluated downbeat scores on RWC-POP MIDI dataset.}

\label{tab:results_midi}
\end{table}

\begin{table}
\centering
\begin{tabular}{|c|c|c|c|}
\hline
\textbf{Input}  & \textbf{Time Step}         & \textbf{Method}    & \textbf{Downbeat F1} \\ \hline

\multirow{4}{*}{Audio} & \multirow{2}{*}{Tatum} & Proposed  & 95.80 $\pm$ 9.20   \\ \cline{3-4}   &   &  Supervised \cite{supervised}   & 96.54 $\pm$ 8.61\\ \cline{2-4} & \multirow{2}{*}{Frame} & Madmom \cite{bock2016joint}\tablefootnote{The model was trained on RWC-POP recordings.}  & 94.68 $\pm$ 17.30   \\ \cline{3-4}   &   &  Transformer \cite{zhao2022beat}$^\textrm{\ref{note:jingwei}}$   & 94.48 $\pm$ 13.69 \\ \hline
\end{tabular}
\caption{Evaluated downbeat scores on RWC-POP audio recordings.}

\label{tab:results_audio}
\end{table}

As another demonstration of the proposed method's utility, we evaluate our models on downbeat detection, the task to classify if a beat is a downbeat or not. For 4/4 songs, the downbeats correspond to the level-4 metrical structure ($l=4$) in our proposed method. We assume correct beats are known at inference time and perform a simple peak picking over $p_{i(\geq 4)}$ to classify downbeats. We train supervised baselines using known downbeat labels from the Lakh (MIDI) or osu (audio) dataset with the same TCN architecture. We also include STOA audio downbeat tracking models (we use synthesized audio for MIDI evaluation) for reference, where predicted downbeats are attributed to the nearest ground-truth beats. The results are shown in table\ \ref{tab:results_midi} for MIDI files and table\ \ref{tab:results_audio} for audio.

The results show that the self-supervised methods' performances are comparable with supervised counterparts, showing that self-supervision is a strong inductive bias for low-level metrical regularity. The self-supervised methods even surpass the supervised method on MIDI datasets. The main reason is that the inaccuracy of downbeat labels\footnote{The MIDI downbeat annotations are calculated based on tempo and time signature events. 5\% songs have wrong downbeats by an inspection of 100 random songs in the Lakh dataset.} harms the performance of supervised methods.

\section{Conclusion}
\label{sec:conclusion}

In this paper, we propose a novel approach to model hierarchical metrical structures for symbolic music and audio signals. We show that the unsupervised CRF loss helps the model learn a musically meaningful metrical hierarchy, and a consistency loss further boosts the performance for multi-track data.  The work shows the potential to improve other rhythm-related MIR tasks with inductive biases based on metrical regularity.

One limitation is that our model requires beat labels for quantization. One future work is to apply self-supervised learning to audio (or performance MIDI) beat tracking, enabling end-to-end metrical structure analysis. Another direction is to explore semi-supervised settings to further boost the model's performance.

\vfill\pagebreak

\bibliographystyle{IEEEbib}
\bibliography{refs}

\end{document}